\documentclass{article}
\usepackage{spconf,amsmath,graphicx}
\usepackage{svg,comment}
\usepackage{booktabs}
\usepackage{enumitem}

\title{Unifying the Discrete and Continuous Emotion labels for Speech Emotion Recognition }
\name{
Roshan Sharma$^1$*,Hira Dhamyal$^2$*, Bhiksha Raj$^2$, Rita Singh$^2$
\thanks{*These authors have equal contribution.}}

\address{$^1$Electrical and Computer Engineering, $^2$Language Technologies Institute \\ Carnegie Mellon University 
}
\begin{document}
\ninept
\maketitle
\begin{abstract}
Traditionally, in paralinguistic analysis for emotion detection from
speech, emotions have been identified with \textit{discrete} or
\textit{dimensional} (continuous-valued) labels. Accordingly, models
that have been proposed for emotion detection use one or the other of
these label types.
However, psychologists like Russell and  Plutchik have proposed
theories and models that unite these views, maintaining that these
representations have shared and complementary information. This paper
is an attempt to validate these viewpoints computationally. To this
end, we propose a model to jointly predict continuous and discrete
emotional attributes and show how the relationship between these can
be utilized to improve the robustness and performance of emotion
recognition tasks. Our approach comprises multi-task and hierarchical
multi-task learning frameworks that jointly model the relationships
between continuous-valued and discrete emotion labels. Experimental
results on two widely used datasets (IEMOCAP and MSPPodcast) for
speech-based emotion recognition show that our model results in
statistically significant improvements in performance over
strong baselines with non-unified approaches.
We also demonstrate that using one type of label (discrete or
continuous-valued) for training improves recognition
performance in tasks that use the other type of label. Experimental
results and reasoning for this approach (called mis-matched training
approach) are also presented.
\end{abstract}
\section{Introduction}
\label{sec:intro}
% Emotion recognition can be performed using multiple modalities\cite{https://doi.org/10.48550/arxiv.2202.08974} including speech, text and video. 
% HOW EMOTION IS REPRESENTED. 

% For this database,
% two of the most popular assessment schemes were used: discrete categorical based annotations (i.e., labels such as happiness, anger, and
% sadness), and continuous attribute based annotations (i.e., activation,
% valence and dominance)

% Speech Emotion Recognition (SER) is a challenging problem even with the current advancements in deep learning.

Emotion is a complex entity which can be modelled in different ways. When defined as a motor response to stimuli, it can be viewed as belonging to a discrete set; whereas when considered as a subjective feeling, it can be expressed as a continuous vector in multiple dimensions~\cite{borod2000neuropsychology}. Though these two definitions of emotion appear to be different, they have shared and complementary information. Psychological models like Plutchik's wheel of emotions~\cite{plutchik1980general} or Russell's ``arousal-valence'' space~\cite{russell1980circumplex} represent discrete emotions on and within a circle on a continuous ``arousal-valence'' plane, demonstrating that the values of ``arousal'' and ``valence'' are correlated to the perception of discrete emotion. 

%Continuous representations offer more flexibility in representing emotions, while discrete emotions contain more specific information that may be useful for downstream applications. 

Speech Emotion Recognition(SER), which refers to the task of identifying the emotional state of the speaker using speech as input, can be used to predict both discrete or continuous emotions. The most commonly used continuous representation for this task comprises three attributes  -- Valence (V), Arousal (A), and Dominance(D) (from Russell \cite{russell1991culture}). Among these three dimensions, Valence represents the pleasantness of the emotion, Arousal denotes the intensity of it, and Dominance represents the degree of control over a social situation. Similarly, a number of discrete emotions have been identified -- happy, sad, angry etc. Translating between these two forms of emotion representations has been of interest ~\cite{kessler2008simplex,  hoffmann2012mapping, buechel2017flexible, trnka2021mapping}, both with audio and textual input. However conflicting results have been found: while some studies\cite{hoffmann2012mapping} find that there is little correlation between continuous values and discrete labels, others~\cite{trnka2021mapping} find correlations like `anger has higher arousal and lower valence'. Based on the notion that discrete and continuous emotion attributes are likely related, in this paper, we propose to examine such dependency relationships.

In order to investigate the relationship between continuous and discrete attributes, we develop neural network models that jointly predict continuous and discrete emotion labels. The motivation behind this is two-fold: (a) holistic models for automatic emotion recognition must be able to capture both generic and specific notions of emotion contained in the continuous and discrete emotion labels, and (b) since discrete and continuous emotion labels may be related to each other, robust and generalizable machine learning models can be built by leveraging this dependency. We propose a multi-task learning framework where continuous and discrete emotion labels are predicted together, but independently of each other. Next, we consider the possibility that knowledge of the discrete emotion might help predict more accurately the continuous emotion and vice versa and hence introduce hierarchical multi-task models that model such a relationship to jointly predict discrete and continuous emotion. 
% In this work, we take a methodical approach to use both discrete and continuous emotion labels. Our baseline model is where both the discrete and continuous labels are predicted independently from the audio.
% Furthermore, we propose a multi-task learning framework where both forms of labels are predicted simultaneously from the input audio. Additionally we propose a hierarchical multi-task learning framework. In this model, we use one type of labels to predict the other. This allows the model to learn separability in one representation of emotion based on the other.

We further demonstrate that these continuous and discrete labels need not necessarily be manually annotated within the same corpus to be improve recognition performance. Of the prevailing annotated datasets for emotion recognition, very few are annotated for both continuous and discrete attributes such as MSPPodcast \cite{lotfian2017building}, and IEMOCAP\cite{BussoBLKMKCLN08IEMOCAP} while others like MELD\cite{poria-etal-2019-meld} are annotated for discrete attributes only. We demonstrate that even if the model is trained on continuous emotion labels from MSPPodcast and discrete labels from IEMOCAP, the proposed Hierarchical Multi-Task approach improves performance and generalizability. This implies that emotion recognition datasets can be trained jointly on multiple corpora with different labels. In summary, this paper makes the following contributions:
\begin{enumerate}[topsep=0pt,itemsep=-1ex,partopsep=1ex,parsep=1ex,leftmargin=*]
    \item We propose a multi-task learning framework that jointly predicts the continuous and discrete labels from speech 
    \item We extend this framework to model hierarchical dependencies, where knowledge of discrete attributes aids continuous prediction and vice versa.
    \item We demonstrate that the proposed approach can be used when the mis-matched labels (continuous and discrete) are drawn from different datasets.
\end{enumerate}

\section{Proposed Approach}
\label{sec:approach}

\begin{figure}
    \centering
    % \includesvg[width=0.45\linewidth,height=0.4\linewidth]{images/discrete_cont_emotion_mtl}
     \includegraphics[width=0.3\linewidth,height=0.4\linewidth]{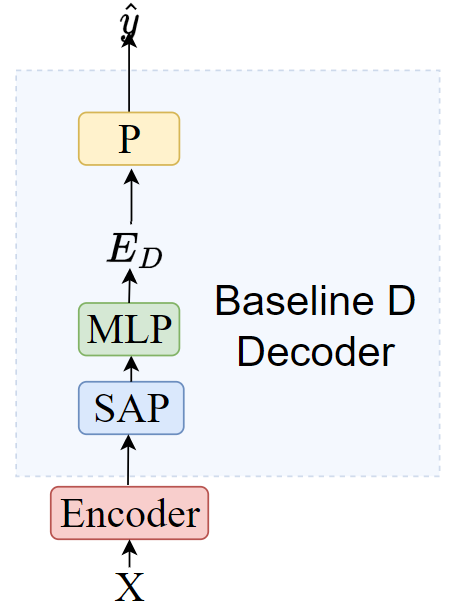}
    \includegraphics[width=0.3\linewidth,height=0.4\linewidth]{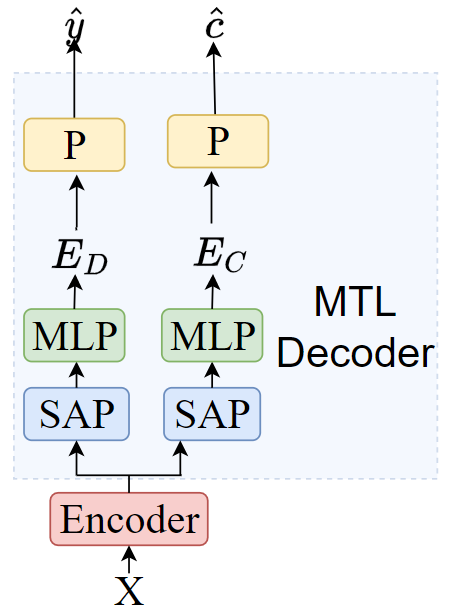}
    \includegraphics[width=0.3\linewidth,height=0.4\linewidth]{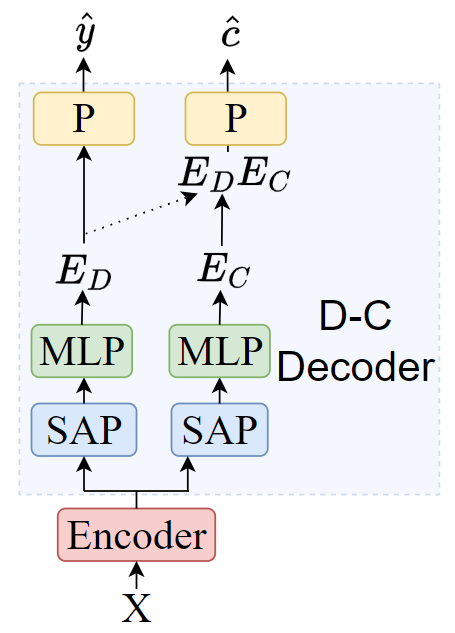}
     \includegraphics[width=0.3\linewidth,height=0.4\linewidth]{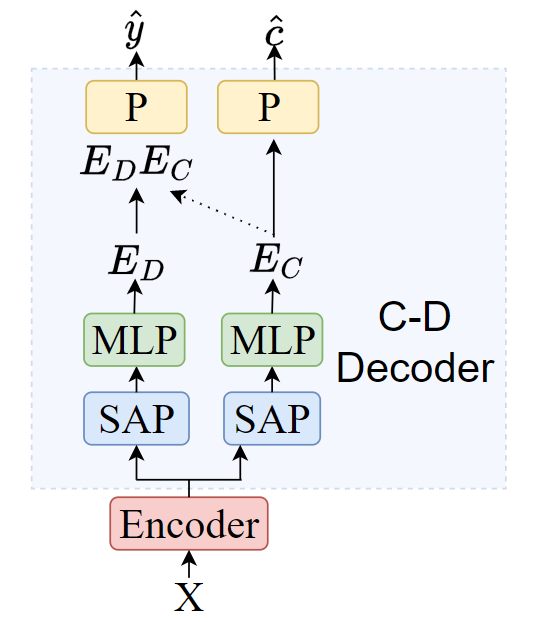}
    \caption{The figure shows model architectures used in this work. The top left shows the Baseline D model. Baseline C model is similar (not shown) with intermediate $E_C$ and final continuous prediction $\hat{c}$. The top middle shows the Multi-task model. Top right and bottom figure show the Hierarchical multi-task model D-C and C-D respectively. SAP stands for self-attention pooling. MLP stands for multi-layer perceptron. $\hat{y}$ is the discrete label prediction whereas $\hat{c}$ is the 3-dimensional continuous emotion prediction.}
    \label{fig:mtl_hmtl}
\end{figure}
% To perform emotion recognition, models take in a sequence of speech frames, and produce a discrete label or a series of 3 continuous predictions.  In this section, we first review the dimensional and discrete models for speech emotion recognition. Then we introduce the proposed multi-task learning method.  

% In this section, we describe the baseline architecture for the continuous and discrete models. Then we introduce the proposed multi-task and hierarchical multi-task structures that model dependency relationships across continuous and discrete emotions. 

\subsection{Encoder Decoder Architecture}
To perform SER on input speech, we use an encoder-decoder architecture for the task. The encoder takes as input a sequence of $N$ content based speech features $X=[x_1,x_2,\cdots x_N]$, and produces as its output a sequence of hidden representations $H = [h_1,h_2,....h_M]$.
These hidden representations are the input to a decoder that either predicts continuous emotion $\hat{c}$, discrete emotion $\hat{y}$, or both.

% Following this, the output is passed to a decoder.

The decoder comprises a temporal self-attentive pooling layer, followed by a Multi-layer Perceptron (MLP) that extracts task specific embeddings, i.e., $E_D$ for discrete prediction and $E_C$ for continuous prediction. These embeddings are then passed through the final classification layer that maps to 3 dimensions for continuous prediction (corresponding to V,A,D), or 5 dimensions for discrete prediction (corresponding to the number of discrete emotion classes).  The continuous and discrete baseline models use a similar neural network architecture, except for the number of output neurons. 

Each of the models we describe below, baseline models that perform either continuous or discrete prediction, and multi-task or hierarchical multi-task models that predict both discrete and continuous emotion, share the same encoder structure but with slightly different decoder architectures.

% , where the output is denoted as $E_D$, which is used for discrete prediction or $E_CT$ which is used for continuous prediction. 

% The encoder is a light-weight module with 4x convolutional subsampling followed by four conformer blocks- 

% extracted from a large-scale self-supervised pre-trained \texttt{HUBERT-large} model that was trained on 60,000h of audiobooks, and uses conformer blocks to aggregate  

\subsection{Baseline Models}
\noindent \textbf{Baseline Discrete Model}\\
To \emph{independently} predict the discrete attributes from input  $X$, our encoder-decoder model termed as \textbf{Baseline D} uses a decoder with discrete self-attentive pooling, MLP which generates the discrete embedding $E_D$, and a discrete classification layer that produces the discrete label $\hat{y}$ as its output. The model is shown in Fig. \ref{fig:mtl_hmtl} (top lefT).

% we build the following baseline model - . 
% Given $X$, the encoder transforms this input sequence into a sequence of hidden representations $H$. 
% Following a sequence of projection and pooling layers on the hidden representation, the model produces a 

% Mathematically, we have 

% \begin{equation}
% H = \text{Encoder}(X) \label{eqn_enc_disc}
% \end{equation}

% \begin{equation}
% P = Pool(H) 
% \label{eqn_pool_disc}
% \end{equation}
    
% \begin{equation}
% \hat{y} = Projection(P)     \label{eqn_projn_disc}
% \end{equation}

Given the true discrete label $y$, the model is optimized using the multi-class cross entropy criterion. For a dataset with $A$ utterances, the cross-entropy loss is computed as:
\begin{equation}
\mathcal{L_\text{disc}} = -\sum_{i=1}^A \sum_{c=1}^K y_{i,c}\log(\hat{y}_{i,c})   
\label{eqn_loss_disc}
\end{equation}

% \subsection{Continuous Emotion Recognition}

\noindent \textbf{Baseline Continuous Model}\\
To predict the continuous attribute labels \emph{independently} from the input speech, our encoder-decoder model, called \textbf{Baseline C} uses a continuous emotion pooling layer and MLP to predict the continuous emotion $\hat{c} =[\hat{c}_{Val},\hat{c}_{Aro},\hat{c}_{Dom}]$.

%. The setting is similar to the previous model. The encoder transforms the input $X$ into hidden representations $H$, which are then used to predict  a three dimensional vector . 
% Using the same definitions as above, instead of predicting the discrete label $\hat{y}$, the model now predicts a three dimensional vector $\hat{c} =[\hat{c}_{Val},\hat{c}_{Aro},\hat{c}_{Dom}]$, where 
The variables $\hat{c}_{Val},\hat{c}_{Aro},$ and $\hat{c}_{Dom}$ correspond to the valence, arousal and dominance predictions respectively. 
% Similar to the discrete case above, let the input to the model is frame sequence $X$ of length $N$. The encoder takes this input and transforms it into hidden representation $H$. Following sequence of pooling and projection layers,

% Given the i-th utterance, the continuous target model produces a continuous vector $\hat{C}^i =[\hat{c}^i_{Val},\hat{c}^i_{Aro},\hat{c}^i_{Dom}]$ of three values, representing valence, arousal and dominance as its predicted output.  
%   Equations \eqref{eqn_enc_disc} and \eqref{eqn_pool_disc} hold, while Equation \ref{eqn_projn_disc} is modified to the formulation in Equation \ref{eqn_projn_cont}.

% \begin{equation}
% \hat{c_\text{Val}},\hat{c_\text{Aro}},\hat{c_\text{Dom}} = Projection(P) \label{eqn_projn_cont}
% \end{equation}
% From \cite{parthasarathy2017jointly}, it is known that a multi-task framework suits the joint prediction of arousal, valence and dominance, and hence we adopt a multi-task framework for the same. 

This model is optimized using the Concordance Correlation Coefficient (CCC) Loss. Given the prediction $\hat{c}$ and the ground truth $c$, the CCC loss is defined as shown in Equation \ref{eqn_cccloss}.
\begin{equation}
    \mathcal{L_\text{ccc}} = 1 - \dfrac{2s_{c\hat{c}}}{s_{c}^2 + s_{\hat{c}}^2 + (\bar{c} -\bar{\hat{c}})^2} \label{eqn_cccloss}
\end{equation} 
where  $s_{c\hat{c}},s_{c}^2,s_{\hat{c}}^2,\bar{c},\bar{\hat{c}}$ represent the covariance between the ground truth and prediction, variance of the ground truth, variance of the prediction, mean of the groundtruth and mean of the prediction respectively.The resulting loss for continuous emotion recognition can be written as the sum of the CCC losses for valence, arousal and dominance prediction.
% \begin{equation}
%     \mathcal{L_\text{cont}} = \mathcal{L_\text{ccc}}^{Val} + \mathcal{L_\text{ccc}}^{Dom} +\mathcal{L_\text{ccc}}^{Aro}
% \end{equation}

% Typically, unit weights are used for the valence, arousal and dominance terms in the loss formulation. 

% \subsection{Converting across Dimensional and Discrete Representations}
\subsection{Multi-Task Model}

Since continuous and discrete representations carry information about the same emotion, we are interested in understanding if \emph{jointly} predicting continuous and discrete attributes improves model performance. To do this, we use a multi-task architecture to predict both the discrete and continuous emotion attributes simultaneously. Since we seek to find the commonality between the discrete and continuous representations, in the multi-task architecture, we elect to utilize a shared speech encoder that transforms the input speech into hidden representations $H$ that contains the information that pertains to both the continuous and discrete emotion attributes,i.e., the shared information. 
% However the model consists of two separate branches which perform the two attribute predictions, where the encoder is shared among the two.

% independently of each other.

%discrete emotions are a good intermediate to attain better continuous attribute scores or vice versa.
% Since dimensional attributes can be reduced to discrete values using a threshold based mapping, 

% this transformation, as in Equation \eqref{eqn_mapping_ct_dt} 
% \begin{equation}
% \label{eqn_mapping_ct_dt}
%     \hat{c_\text{Val}},\hat{c_\text{Aro}},\hat{c_\text{Dom}} = Projection(\hat{y})
% \end{equation}

% We employ the same learnable projection to obtain continuous representations from discrete outputs (see Equation \eqref{eqn_mapping_dt_ct}). 
% \begin{equation}
% \label{eqn_mapping_dt_ct}
%     \hat{y} = Projection(\hat{c_\text{Val}},\hat{c_\text{Aro}},\hat{c_\text{Dom}})
% \end{equation}

% \subsection{Multi-task Modelling for Dimensional and Discrete Emotion Recognition}

%for jointly learning to predict discrete and continuous emotion attributes.
The \textbf{Multi-task C,D} model predicts both discrete label $\hat{y}$ and continuous vector 
$\hat{c} =[\hat{c}_{Val},\hat{c}_{Aro},\hat{c}_{Dom}]$ for every utterance. 

The architecture of the model is shown in Fig. \ref{fig:mtl_hmtl} (left). The model uses a common encoder, but within the decoder- two parallel branches are used to learn task specific pooling and MLP parameters. The discrete branch (left) generates a discrete embedding $E_D$, which is then used to predict the discrete emotion $\hat{y}$. Similarly, the continuous branch predicts a continuous embedding $E_C$, which is used to predict the continuous emotion $\hat{c}$. 
% The encoder maps input speech frames $X$ to a hidden representation which is passed through self-attention pooling layer (SAP), followed by a linear layer. 
% One part of the model projects into $E_D$ which is used to predict the discrete emotion $\hat{y}$. The other part of the model projects to $E_C$, which is used to predict the continuous emotion $\hat{c}$. 
The model is optimized with the total loss as shown in Equation \eqref{eqn_mtl_loss}. 
\begin{equation}
    \mathcal{L_\text{total}} = \alpha(\mathcal{L_\text{ccc}}^{Val} + \mathcal{L_\text{ccc}}^{Dom} +\mathcal{L_\text{ccc}}^{Aro}) + \beta\mathcal{L_\text{disc}}
    \label{eqn_mtl_loss}
\end{equation}

where the values of $\alpha$ and $\beta$ are set to 1 empirically.

\subsection{Hierarchical Multi-Task Model}
Multi-task modelling described in the previous section seeks to utilize the shared information between discrete and dimensional attributes to predict them jointly. However, it doesn't assume any direct dependency between the discrete and continuous emotion attributes. Based on the hypothesis that knowledge of discrete emotion attributes would help improve continuous attribute predictions and vice-versa, we develop hierarchical multi-task models.

In this model, the continuous and discrete emotion prediction branches in the decoder are used to generate the continuous and discrete embeddings $E_C$ and $E_D$ respectively, as in the multi-task formulation. However, here, the predicted continuous emotion, i.e., $\hat{c}$ is not computed solely based on $E_C$, but also on $E_D$.  In other words, the discrete emotion embedding is used in conjunction with the continuous embeddding to predict the continuous emotion. We term this our \textbf{Hierarchical D-C} model since the discrete embedding is used as auxiliary input to predict the continuous emotion. Similarly, one can define a \textbf{Hierarchical C-D} model where the continuous emotion embedding is used as auxiliary input to predict the discrete emotion. 

The \textbf{Hierarchical D-C} model is shown in Fig. \ref{fig:mtl_hmtl} (right). To perform continuous emotion prediction with the help of the discrete predictions, 
 discrete emotion embedding $E_D$, (which is used to predict the discrete emotion $\hat{y}$) is concatenated with the continuous embedding $E_C$. This concatenated embedding $[E_{D}E_{C}]$ is then used to predict the continuous emotion, $\hat{c}$. Similarly, in the \textbf{Hierarchical C-D} model, in order to utilize continuous attributes to predict discrete attributes, $E_C$ is concatenated with $E_D$ and used to predict $\hat{y}$. 

\section{Experiments}
\label{sec:results}
\subsection{Data}
Our experiments are conducted on the MSPPodcast\cite{Lotfian_2019_3} and IEMOCAP \cite{busso2008iemocap}. MSP Podcast is the largest human labelled emotion dataset with 29,965 training examples and 10,013 test utterances. It is comprised of segments of speech from podcasts, which means that the speech is expressive and the acoustic environment less constrained by background noise and other interference. The MSPPodcast data has labels for three continuous dimensions - valence, arousal and dominance, and for multiple discrete emotions - of which we use five : neutral, angry, happy, sad, and disgust. In this work, we report our results on the balanced test1 evaluation set with 30 male and 30 female speakers. 
The second dataset we use is IEMOCAP which is a 20 hour dataset, with annotations on acted emotion. It comprises of 10 different speakers, 5 male and 5 female. It is labelled for three continuous dimensions - valence, arousal and dominance, and for 9 discrete emotions out of which we utilise the ones that overlap with our MSPPodcast set, i.e., neutral, angry, happy, sad, and disgust.

\subsection{Model Hyperparameters}
All our models are built using the ESPNet\cite{watanabe2018espnet} toolkit, and will be publicly released to encourage further research.
% Our code will be made publicly available to encourage further research in this direction. 
% Self-supervised representation learning models are among the start of the art for many tasks including speech emotion recognition, and hence 
We use \textit{HUBERT-large} \cite{hsu2021hubert} embeddings as the input features and a 4 layer conformer encoder with 64 hidden units for our models.  We freeze the \texttt{HUBERT-large} frontend, and train the conformer and pooling decoder parameters. We employ separate self-attentive pooling\cite{safari2020self} layers for continuous and discrete emotion prediction so the model can learn which frames to focus on in order to make emotion predictions. Our decoders consist of linear projections that map from the encoded output dimension of 768 to 64, 32 and consequently the output sizes of 3 for the continuous prediction and 5 for the discrete prediction.  We use a ReLU activation to ensure that the predicted continuous attributes are strictly positive and use LeakyReLU elsewhere. We use a dropout of 0.2 in the decoder. Our models are trained with the Adam optimizer, and a peak learning rate of 1e-3 for 15,000 warmup steps.

%model without ASR finetuning, and freeze the parameters in the CNN feature extractor while training all our models.
%HUBERT\cite{hsu2021hubert} is an iterative estimator with masked acoustic frame classification as the proxy task. Subsequent iterations produce different classification targets as clustering is performed on different feature sets including MFCCs and intermediate layer activations from previous iterations. The base HUBERT model comprises CNN feature extractors followed by 12 BERT-like transformer layers.

\subsection{Evaluation Metrics}
Discrete emotion recognition is evaluated using F1 or accuracy. For continuous prediction, we compute the Concordance Correlation Coefficient for each of the attributes arousal, valence and dominance, and consequently their mean. Equation \eqref{eqn_ccc_metric} shows how the CCC is computed given prediction $\hat{c}$, and ground truth $c$, where $s_{c\hat{c}},s_{c}^2,s_{\hat{c}}^2,\bar{c},\bar{\hat{c}}$ represent the covariance between the ground truth and prediction, variance of the ground truth, variance of the prediction, mean of the groundtruth and mean of the prediction respectively.  

\begin{equation}
    \mathcal{\text{CCC}} = \dfrac{2s_{c\hat{c}}}{s_{c}^2 + s_{\hat{c}}^2 + (\bar{c} -\bar{\hat{c}})^2} \label{eqn_ccc_metric}
\end{equation}

% \begin{table}[tbh!]
% \caption{Emotion Recognition Results on IEMOCAP using 5-fold cross-validation: Concordance Correlation Coefficient- overall,Valence,Activation and Dominance is reported for continuous emotions, and Unweighted F1 is reported for discrete emotion prediction}
% \resizebox{0.5\textwidth}{!}{%
% \begin{tabular}{l|rlllrr}
% \toprule
% Model Description    & \multicolumn{1}{l}{CCC} & CCC-V & CCC-A & CCC-D &  F1\\ 
% \midrule
% Baseline C                    & 0.406                             & 0.528         & 0.568        & 0.121        & \multicolumn{1}{}{-}           \\
% Baseline D                    & \multicolumn{1}{}{-}             &       -         &       -         &     -           &    0.627                        \\
% Multi-task C,D                & 0.388                            & \textbf{0.542}        & 0.619        & 0.003        &   0.613                         \\
% Hierarchical D-C & \textbf{0.551}                            & 0.396        & \textbf{0.654}        & \textbf{0.602}        &      0.641                    \\
% Hierarchical C-D & 0.529                            & 0.458        & 0.583        & 0.545         &  \textbf{0.648} \\ 
% \bottomrule
% \end{tabular}
% }
% \label{tab:iemocap_results}
% \end{table}

\begin{table}[tbh!]
\caption{Emotion Recognition Results on IEMOCAP using 5-fold cross-validation: Concordance Correlation Coefficient- overall,Valence,Activation and Dominance is reported for continuous emotions, and Unweighted accuracy is reported for discrete emotion prediction}
\resizebox{0.5\textwidth}{!}{%
\begin{tabular}{l|ccccc}
\toprule
Model Description    & \multicolumn{1}{l}{CCC} & CCC-V & CCC-A & CCC-D &  Acc\\ 
\midrule
Baseline C                    & 0.580                             & 0.548         & 0.606        & 0.566        & \multicolumn{1}{c}{-}           \\
Baseline D                    & \multicolumn{1}{c}{-}             &       -         &       -         &     -           &    0.737                       \\
Multi-task C,D                & 0.603                            & 0.571        & 0.669        & 0.567        &   0.723                         \\
Hierarchical D-C & \textbf{0.667}                           & \textbf{0.660}        & \textbf{0.717}        & \textbf{0.625}        &      0.744                    \\
Hierarchical C-D & 0.648                            & 0.651        & 0.694        & 0.599         &  \textbf{0.749} \\ 
\bottomrule
\end{tabular}
}
\label{tab:iemocap_results}
\end{table}

% Please add the following required packages to your document preamble:
% \usepackage{graphicx}
\begin{table*}[tbh!]
\caption{Results on MSPPodcast and IEMOCAP: The second two columns represent the attributes used in training from the IEMOCAP and MSPPodcast datasets respectively. CCC is reported on IEMOCAP Session05 and on the test1 evaluation sets for MSPPodcast respectively. F1 is reported on the test1 evaluation set for MSPPodcast and unweighted accuracy on Session05 for IEMOCAP (we use this because these are standard metrics reported for these datasets).}
% \resizebox{0.5\textwidth}{!}{%
\centering
\begin{tabular}{lcccccc}
\toprule
ID & \textbf{IEMOCAP Train} & \textbf{MSPPodcast Train} & \multicolumn{1}{c}{\textbf{IEMOCAP CCC}} & \multicolumn{1}{c}{\textbf{IEMOCAP Acc}} & \multicolumn{1}{c}{\textbf{MSPPodcast CCC}} & \multicolumn{1}{l}{\textbf{MSPPodcast F1}} \\ \midrule
1 & -                & Cont.                & 0.125                             & \multicolumn{1}{c}{-}              & 0.524                             & \multicolumn{1}{c}{-}             \\
2 & Cont.               & -                & 0.327                             & \multicolumn{1}{c}{-}              & 0.140                          & \multicolumn{1}{c}{-}             \\ 
3 & -                & Disc.                & \multicolumn{1}{c}{-}              & 0.459                             & \multicolumn{1}{c}{-}              & 0.325                            \\
4 & Disc.                & -                & \multicolumn{1}{c}{-}              & 0.701                             & \multicolumn{1}{c}{-}              & 0.266                            \\ 
5 & Cont.                & Cont.                & 0.586                             & \multicolumn{1}{c}{-}              & 0.524                             & \multicolumn{1}{c}{-}             \\ 
6 & Disc.                & Disc.                & \multicolumn{1}{c}{-}              & 0.695                             & \multicolumn{1}{c}{-}              & 0.364                            \\ 
7 & Cont.                & Disc.                & 0.625                             & 0.470                             & 0.189                              & 0.302                            \\ 
8 & Disc.                & Cont.                & 0.218                              & 0.705                             & 0.497                             & 0.283                            \\ 
\bottomrule
\end{tabular}%
% }
\label{tab:msp_iemocap_results}
\end{table*}

\subsection{Results on IEMOCAP}

Table \ref{tab:iemocap_results} shows our experimental results on the IEMOCAP dataset, where all results are computed using standard 5-fold cross validation~\cite{Morias22}. We observe that multi-task training improves CCC by 0.02 over the continuous baseline, however, it does not outperform the discrete baseline. We contend that this is because both of these predictions are being generated independent of the other.

% As a reminder, continuous and discrete emotion prediction baselines are built using HUBERT features and the conformer architecture. We note that our baselines are comparable to other works~\cite{Morias22} (which obtains an un-weighted accuracy of 0.7345 using HuBERT features).  

The proposed hierarchical models outperform the baselines and multi-task model on discrete and continuous emotion prediction. Specifically the hierarchical model that uses discrete attributes to aid the prediction of continuous attributes, i.e., Hierarchical D-C outperforms the multi-task model and the continuous baselines on CCC - gaining $0.087$ absolute CCC. This is perhaps because knowledge of the discrete emotion aids in the prediction of dominance and arousal. We also observe an absolute $0.8\%$ improvement on accuracy in Hierarchical D-C. The hierarchical model that uses continuous emotions to help predict discrete attributes, i.e., Hierarchical C-D outperforms the other 3 models on accuracy. This reaffirms our hypothesis that knowledge of discrete and continuous emotion attributes can be used to improve performance on continuous and discrete emotion prediction respectively. We note that the Hierarchical C-D model improves the recognition of discrete emotion attributes compared to the multi-task model and the continuous baseline.

\subsection{Results on MSPPodcast}

\begin{table}[tbh!]
\caption{Results on MSPPodcast: Concordance Correlation Coefficient(CCC) - overall,Valence,Activation and Dominance is reported on the test1 evaluation set. For discrete emotions, unweighted F1 is reported.}
\resizebox{0.5\textwidth}{!}{%
\begin{tabular}{l|ccccc}
\toprule
Model Description             & \multicolumn{1}{c}{CCC} & \multicolumn{1}{c}{CCC-V} & \multicolumn{1}{c}{CCC-A} & \multicolumn{1}{c}{CCC-D} & \multicolumn{1}{c}{F1} \\ 
\midrule
Baseline C                    & 0.593                   & \textbf{0.597}                     & 0.646                     & 0.538                     & \multicolumn{1}{c}{-}   \\
Baseline D                    & \multicolumn{1}{c}{-}   & \multicolumn{1}{c}{-}     & \multicolumn{1}{c}{-}     & \multicolumn{1}{c}{-}     & 0.368                   \\
Multi-task C,D                & 0.587                   & 0.591                     & 0.637                     & 0.533                     & 0.393                  \\
Hierarchical D-C & \textbf{0.617}                   & 0.588                     & \textbf{0.675}                     & \textbf{0.584}                     & 0.404                  \\
Hierarchical C-D & 0.605                   & 0.554                     & 0.661                      & 0.569                     & \textbf{0.411}     \\ 
\bottomrule
\end{tabular}%
}
\label{tab:msp_results}
\end{table}
Table \ref{tab:msp_results} reports the results of experiments on MSPPodcast. 
The continuous and discrete emotion prediction baselines obtain comparable scores to state-of-the-art approaches\cite{Srinivasan22}.

Multi-task training to jointly predict continuous and discrete attributes improves F1 score over the discrete baseline while slightly degrading continuous emotion prediction performance. As with the IEMOCAP dataset, the proposed hierarchical models outperform the baselines and multi-task prediction models. Specifically, the hierarchical D-C model improves CCC on continuous attribute prediction over the multi-task and baseline continuous models. 
Hierarchical multi-task learning helps learn useful intermediate representations~\cite{Sanabria18}. Therefore in the hierarchical D-C model, the continuous prediction task helps learn better discrete emotion representations, thereby improving discrete prediction as well with respect to the baseline. The Hierarchical C-D model also improves prediction of discrete emotion over the other 3 models. This model also outperforms the baseline and multi-task models on continuous emotion prediction, with gains arising from improved prediction of arousal and dominance. 

% In Hierarchical Multi-Task Learning, the tasks at the output layer contribute to the task at lower layers during the backward pass. Effectively, in the D-C model, gradients from the prediction of continuous emotion contribute to the updates in parameters pertraining to discrete emotion prediction. Similarly, in the C-D model, gradients from the prediction of discrete emotion contribute to the update of parameters that pertain to continuous emotion prediction.
% Since converting continuous to discrete labels is a lossy process, going from discrete to continuous in the D-C is an inexact process. We believe this explains why D-C model is more effective than in the C-D model on the lower task , i.e., discrete emotion prediction.   
% is the reasons for our observation of we observe that the regularization resulting from the D-C model is more effective than in the C-D model on the lower task , i.e., discrete emotion prediction.   
% Since the process of converting continuous attributes to discrete attributes is more direct and easier to perform than the conversion from discrete to continuous attributes, we observe that the regularization resulting from the D-C model is more effective than in the C-D model on the lower task , i.e., discrete emotion prediction.   

\subsection{Combining IEMOCAP and MSPPodcast}
In this section, we attempt to use discrete and continuous labels from different datasets and analyse gains when training models on multiple datasets with matched (e.g. discrete-discrete) and mis-matched (e.g. continuous-discrete) labels. 
We perform all experiments using a subset of the MSPPodcast data chosen randomly such that the resulting size is the same as that of the IEMOCAP training data. This is done in order to be able to make fair comparisons on transferrability of representations. 
Table \ref{tab:msp_iemocap_results} summarizes the results of our experiment on transferring labels across datasets. 

\noindent\textbf{IEMOCAP Discrete Prediction}: Consider rows with ID 4,6,8 from Table \ref{tab:msp_iemocap_results}- we observe that the best unweighted accuracy on the IEMOCAP test set is obtained when the model is trained on IEMOCAP discrete labels and MSPPodcast continuous labels. 

\noindent\textbf{IEMOCAP Continuous Prediction}: Comparing rows 2,5,7 from Table \ref{tab:msp_iemocap_results}, we observe that the best CCC on IEMOCAP is achieved when IEMOCAP continuous labels and MSPPodcast discrete labels are both included in the training. 

\noindent\textbf{MSPPodcast Discrete Prediction}: From rows 3,6,7 in Table \ref{tab:msp_iemocap_results}, we observe that the best F-1 score is obtained on  MSPPodcast when discrete labels from both datasets are used, i.e., for row 6 model. 

\noindent\textbf{MSPPodcast Continuous Prediction}: From  the rows 1,5,8 from Table \ref{tab:msp_iemocap_results}, we can conclude that when the MSPPodcast continuous labels are used alone, the best performance is achieved. 

In conclusion, we observe that using mis-matched labels from the MSPPodcast data for training improves performance on discrete and continuous emotion prediction for IEMOCAP. We also note that though discrete labels from IEMOCAP are transferrable and improve performance on MSPPodcast, the continuous labels from IEMOCAP do not seem to provide gains on MSPPodcast. 

Therefore MSPPodcast discrete and continuous representations help improve performance on the mismatched label of IEMOCAP, while such transferred representations from IEMOCAP do not significantly impact MSPPodcast predictions. We believe this is in part because of the difference in the nature of annotations across the IEMOCAP and MSPPodcast datasets. For example, from our analysis we observe that anger is assigned a lower arousal and higher valence than neutral in IEMOCAP while in MSPPodcast, anger is assigned a higher arousal and lower valence than neutral. Furthermore, the inter-annotator agreement for continuous values in MSPPodcast are much lower in the IEMOCAP, which makes it challenging to obtain gains in MSPPodcast continuous predictions using transferred representations from IEMOCAP.

\section{Conclusion and Future Work}
\label{sec:conclusion}

Emotion Recognition remains a challenging task. Different datasets are labelled with varying annotation labels, making it challenging to train large scale emotion models utilising all the datasets. In this paper, we introduce hierarchical multi-task learning models that predict discrete or continuous labels by using continuous or discrete labels respectively. 

With our method, we obtain absolute improvements of 1.2 \% Accuracy and 4.3 points F-1 for discrete prediction on IEMOCAP and MSPPodcast respectively. On continuous labels, we improve 0.09 CCC for IEMOCAP, and 0.025 CCC in MSPPodcast. Furthermore, we also combine IEMOCAP and MSPPodcast with mismatched emotion annotations, and show that mis-matched labels from MSPPodcast help performance on IEMOCAP. 
% mutual information in discrete labels and continuous attributes for speech emotion recognition. The proposed approach would make it easier to train jointly on multiple corpora despite having different annotations.

% We then propose to use speaker recognition as an auxiliary task within this multitask framework , which boosts the performance on valence and discrete classification, while retaining comparable performance on arousal and dominance. We introduce two sets of hierarchical structures- one which directly maps from discrete to dimensional variables, and another which uses pooled features in conjunction with discrete variables to predict dimensional variables. The proposed hierarchical structure improves performance on both tasks, 

% \section{Acknowledgement}
% This material is based upon work supported by the Defence Science and Technology Agency, Singapore under contract number A025959. Its content does not reflect the position or policy of DSTA and no official endorsement should be inferred. This work used the Bridges system ~\cite{nystrom2015bridges}, which is supported by NSF award number ACI-1445606, at the Pittsburgh Supercomputing Center (PSC).

% References should be produced using the bibtex program from suitable
% BiBTeX files (here: strings, refs, manuals). The IEEEbib.bst bibliography
% style file from IEEE produces unsorted bibliography list.
% -------------------------------------------------------------------------
\newpage
\bibliographystyle{IEEEbib}
\bibliography{strings,refs}

\end{document}